# CLUSTERING FEATURES OF $^{14}N$ IN RELATIVISTIC MULTIFRAGMENTATION PROCESS

*T.V. Shchedrina[1], P.I. Zarubin[1]*

[1] *Joint Institute for Nuclear Research, Russia*

Abstract: Progress in the study of the peripheral nuclear interactions in $^{14}N$ dissociation at *2.1A GeV* in nuclear emulsion is outlined. The leading role of the multiple fragmentations in the most peripheral nucleus interactions is discussed. The production of unusual states, for which a regrouping of nucleons beyond the α-particle clustering is needed, is identified for this channel.

## 1. Introduction

The BECQUEREL project deals with the investigation of the cluster degrees of freedom in light nuclei [1] which is based on the study of the peripheral interactions of nuclei at energy higher than *1A GeV* in a nuclear emulsion exposed to JINR Nuclotron beams. This method has been used before for the study of α-particle clustering in the dissociation of relativistic $^{14}N$ [3], $^{12}C$ [4] and $^{16}O$ [5] nuclei. Recent results are associated with lighter $^{10}B$ [6], $^{11}B$ [7], $^{9}Be$ [8,9], $^{7}Be$ [8,10] and $^{8}B$ [8,11] nuclei in which nucleon clustering to the lightest $^{2,3}H$ and $^{3}He$ nuclei plays a noticeable role.

The present paper is devoted to the progress in a detailed study of nucleon clustering in the $^{14}N$ nucleus dissociation [8,12]. The main interest in the study of $^{14}N$ was to enlarge the ideas about the role of the α-cluster degrees of freedom in this nucleus as an intermediate one between $^{12}C$ and $^{16}O$. At the same time, the odd-odd $^{14}N$ nucleus gives a convenient possibility of studying the production of nuclear configurations beyond the α-clustering pattern.

Peripheral collisions proceed with a minimum overlapping of the colliding nucleus densities which results in the conservation in a narrow angular cone of the total charge and the atomic weight of an incident nucleus by the produced relativistic fragment systems. Their various configurations include mainly *H* and *He* isotopes with small relative velocities, as well as single heavier nuclei. The most peripheral interactions on heavy *Ag* and *Br* nuclei entering the emulsion compound are not accompanied by other reaction products except the fragments of the nucleus in question - these are the so-called "white" stars [13]. The examples of such interactions are given in fig. 1 and 2 [1]. Their selection makes it possible to count on a reliable revealing of the main cluster features of the nuclei studied, as well as on the discovery of unusual configurations with participation of the largest number of nucleons.

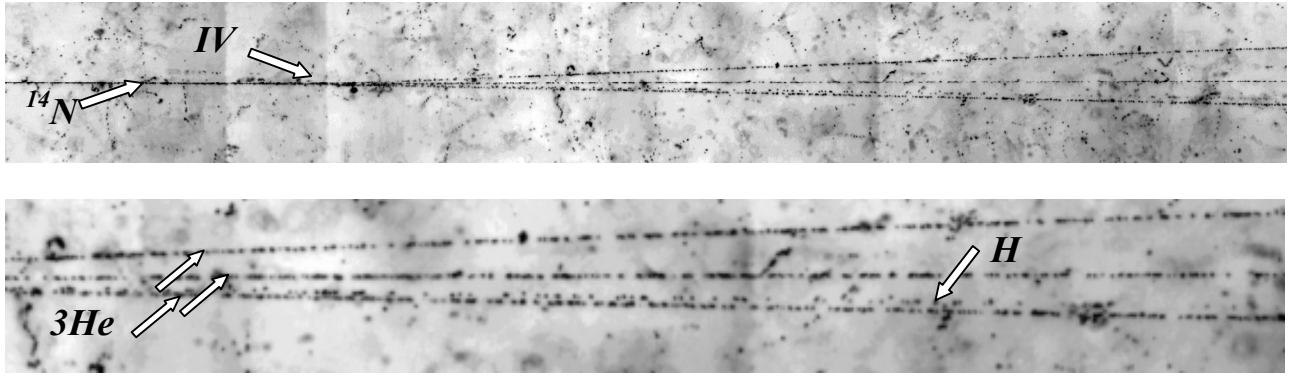

***Fig.1.*** Microphotograph of the nucleus fragmentation $^{14}N \rightarrow 3He + H$ at *2.1A GeV* in peripheral interaction in emulsion. The upper photograph shows the interaction vertex without target fragment production and the fragment jet. In displacing along the fragment jet (lower photo) it is possible to see three double-charged (*He*) and one single-charged fragment (*H*).

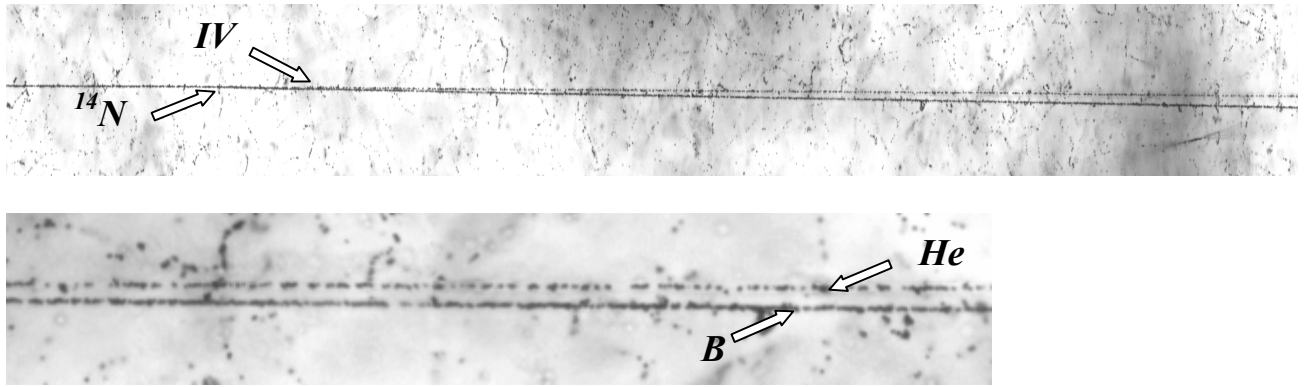

***Fig.2.*** Microphotograph of the nucleus fragmentation $^{14}N \rightarrow {}^{10}B + {}^{4}He$ at *2.1A GeV* in peripheral interaction in emulsion. The upper photograph shows the interaction vertex *IV* and a narrow jet involving two fragments. In displacing along the fragment jet (lower photo) one sees clearly one multi-charged fragment with $Z = 5$ or $^{10}B$ and one double-charged fragment $^{4}He$.

## 2. Possibilities of the method

The relativistic energy of the studied nuclei in combination with the record angular resolution of nuclear emulsion make measurements of the emission angles of relativistic fragments most complete and accurate. The relativistic *H* and *He* charges are easily identified by visual scanning tracks. In order to define the charge of heavier $Z > 2$ particles use is made of the method of counting of $\delta$-electrons ($N_\delta$) per length unit of the track in question. The $Z_{pr}$ charges were measured on the beam nucleus tracks by counting the number of $\delta$-electrons on *3 - 5* cm lengths. The results which illustrate the accuracy of the method are given in fig. 3a. The fraction of the $^{14}N$ nuclei is *81 %*, the remaining *15%* and *4 %* are the accompanying secondary $^{12}C$ and $^{10}B$ nuclei.

The same method was employed to obtain the charge distribution for secondary Z fragments (fig. 3b). An expected change in the distribution compared with fig. 3a is observed. The dependence of the square root $N_\delta$ on the identified $Z_{fr}$ charge is linear enough.

A complete identification of the relativistic H and He fragments is of undoubted interest. In spite of the fact that this method is rather laborious, measurements of the multiple scattering of relativistic fragment tracks makes it possible to identify the H and He isotopes by the product of their momentum values $p$ by the velocity $\beta$ ones - $p\beta c$. The difficulty consists in the following. To define the $p\beta c$ values it is necessary to measure shifts along the track coordinate in more than 100 points. Identification is made under the assumption that the bombarding nucleus fragments approximately keep their initial momentum per nucleon and the velocities equal to the $p_0$ and $\beta_0$ values, and then the fragment mass number is equal to $A_{fr} \approx p\beta c/p_0\beta_0 c$. Because of different technical problems associated with the emulsion layers, as well as through some restrictions related to the angular dispersion of the fragment tracks it turns out to be impossible to carry out these measurements for all the events in a total volume. However such a procedure is realizable for some valuable events. In this case, the interpretation of the event of the relativistic peripheral dissociation of a light nucleus becomes unique as concerns the level of its detailed presentation.

Fig.4 shows the reconstructed spectrum $A_{fr}$ of single-charged and double-charged fragments for the fragmentation channels $^{14}N \to 3He + H$ and $^{14}N \to {}^{10}B + {}^{4}He$. The obtained experimental $A_{fr}$ values for single-charged fragments are described by the sum of two normal distributions. The distribution peaks are located at the $A_f$ values corresponding to hydrogen $^{1}H$ and $^{2}H$ isotopes. There is also one single-charged $A_{fr} = 3$ fragment identified as $^3H$. The $A_{fr}$ area for double-charged fragments is described by the sum of three normal distributions. The approximating distribution maxima for this area are in accordance with the mass numbers of the helium $^3He$, $^4He$ and $^6He$ isotopes. In order to be sure of the fact that it is exactly the $^6He$ isotope that was detected, all the fragments in which this isotope was discovered were identified. One determined the charge and one measured the *pβc* value for each fragment.

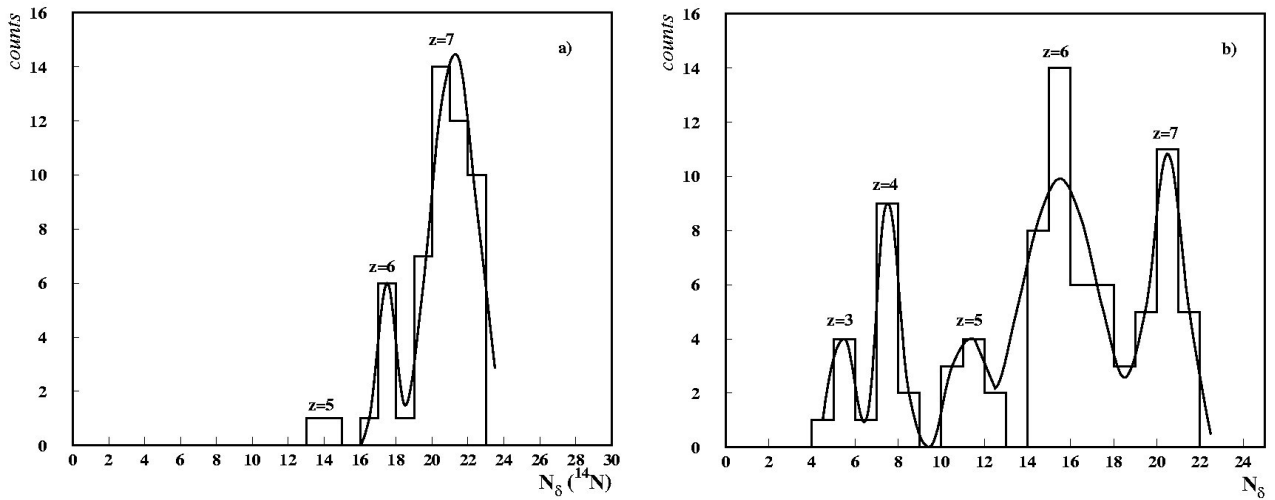

***Fig.3.*** Distribution by the $\delta$-electron number per *1* mm of the track on: a) primary $^{14}N$ nuclei, b) $^{14}N$ nucleus fragments with *3 - 7* charges. The continuous line is the description by the Gauss function sum.

### 3. The major dissociation channel

Nuclear emulsion was exposed to a $^{14}N$ beam of energy of *2.1A GeV*. Unique possibilities of the method have already enabled one to obtain information about the charge topology of the fragments, the distributions over the emission angles and partially about the isotopic compound of relativistic fragments [8, 12].

A systematic search for the interaction vertices using the beam nucleus tracks made it possible to establish that the dissociation channel $^{14}N \rightarrow 3He + H$ is a leading one. It gives a contribution equal to *35 %* both for "white" star statistics and events accompanied by the production of target fragments and mesons. Thus the $^{14}N$ nucleus is displayed as an effective source of *3α*-particle states.

An accelerated viewing enabled one to obtain a total of *132* events in this channel which made it possible to estimate the energy scale of *3α*-particle systems produced in peripheral fragmentation. An invariant estimation of the energy scale of $3\alpha$-system production performed under sufficiently reliable assumptions shows that *80%* of interactions are concentrated in the region below *14 MeV* which corresponds to $^{12}C$ cluster excitations. The contribution of the events $^{14}N \rightarrow {}^{8}Be + \alpha + X \rightarrow 3\alpha + X$ accompanied by an $^{8}Be$ decay from the ground state amounts to about *25%*.

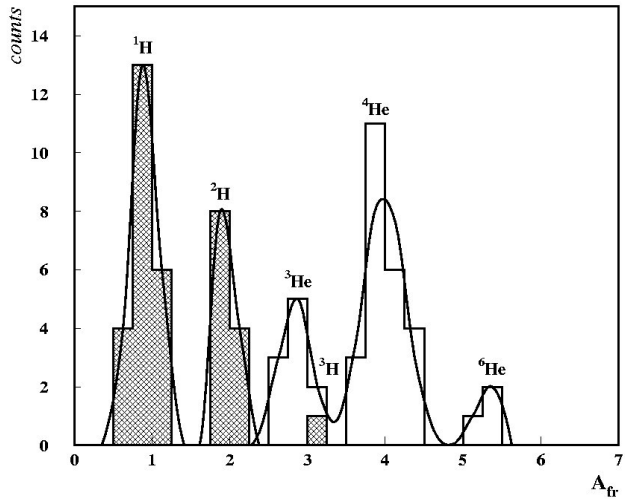

*Fig.4.* Distribution by the mass number $A_{fr} = pβc/p_0β_0c$ (where *pβc* is the experimentally obtained value, $p_0 = 2.86A\ GeV/c$, $β_0 = 0.9$) of single-charged (shaded histogram) and double-charged (non-shaded histogram) fragments for the fragmentation $^{14}N \rightarrow 3He + H$ and $^{14}N \rightarrow C + H$ channels.

The identification of relativistic H nuclei in the channel $^{14}N \rightarrow 3He + H$ points to a noticeable decrease of the deuteron yield with respect to the protons compared with early studied cases of relativistic fragmentation $^{6}Li \rightarrow He + H$ [14] and $^{10}B \rightarrow 2He + H$ [6]. This fact is in accordance with the thresholds of deuteron and proton separation in these nuclei.

The leading role of the multi-particle channel is found to be explicit and, at the same time, unexpected. As was expected initially, the dissociation channel probabilities would have decreased with increasing mass threshold Q. Its value for the channel $^{14}N \rightarrow 3He + H$ is $Q \approx 17$ *MeV*. The largest probability would have corresponded to the channels $^{14}N \rightarrow {}^{13,12}C + {}^{1,2}H$ with noticeably lower values $Q \approx 7.6$ and *10.2* MeV. Really the part of the *C + H* events was found to be rather considerable – 25%. The channel $^{14}N \rightarrow {}^{10}B + {}^{4}He$ ($Q \approx 11.6$ *MeV*) which is reliably observed in emulsion could have been the next one. However the *B + He* fraction was found to be small: no more than *8 %*.

It should be noted that a similar situation was observed when studying "white" stars produced in nuclear emulsion by relativistic $^{10}B$ nuclei [13]. Among them, the three-body channel $^{10}B \rightarrow 2He + H$ with $Q \approx 6$ *MeV* was a leader (*75 %*) while the two-body one $^{10}B \rightarrow {}^{6}Li + {}^{4}He$ with $Q \approx 4.5$ *MeV* made no more than *15%*. Thus for $^{10}B$ and $^{14}N$ nuclei the channel associated with the separation of $^{4}He$ turns out to be suppressed compared with multi-particle dissociation. The probability of the channel $^{14}N \rightarrow {}^{6}Li + 2{}^{4}He$ is also small.

It seems that the leading effect of the multi-particle fragmentation deserves theoretical foundation. It can be due to the nature of multi-particle fragmentation as a phase transition of nucleons from the bound state to the dilute quantum gas produced by the lightest *H* and *He* nuclei. It is quite possible that these processes discussed here for light nuclei underlie also the multi-particle fragmentation (total destruction) of heavier nuclei.

### 4. Rare events

We have succeeded in performing the total identification of fragments for a part of "white" *3He + H* stars. In spite of the fact that it is impossible to present data systematically, it is worth noting the production of unusual states for which nucleon regrouping beyond the alpha bounds and so overcoming of high energy thresholds $Q$ are needed. The facts themselves of their observation can possibly turn out to be useful for understanding the dynamics of relativistic multi-fragmentation. The following channels are identified among these events:

1. Two events $^6He + {}^4He + {}^3He + p$ ($Q \approx 39$ MeV). The total identification makes it possible to estimate the mean transverse momentum transferred to the fragment system $<p_t\ (^6He + {}^4He + {}^3He + {}^1H)> = (431 \pm 43)$ MeV/c. This value can be compared with a small value for the more probable channel $<p_t\ (3^4He + {}^2H)> = (182 \pm 90)$ MeV/c estimated by four totally identified "white" stars. The summary value $p\beta c$ for fragments $^6He + {}^4He + {}^3He + p$ is $(37 \pm 4)$ GeV and $(41 \pm 5)$ GeV for an expected $p\beta c(^{14}N) = 40$ GeV.

2. Two events $^4He + 2^3He + {}^2H$ ($Q \approx 59$ MeV). For this process to go breaking of two α-clusters and emission of a neutron pair are needed.

3. One event $^{11}C + {}^3H$ ($Q \approx 23$ MeV).

"White" stars associated with inelastic charge exchange $^{14}N \rightarrow 3He + 2H$ (9), $^{14}N \rightarrow 3He$ (5), $^{14}N \rightarrow 2He + 2H$ (3) were observed. We stress the reliability of identification of the involved nucleus charges. In this case, of special interest are the events $2He + 2H$. The conditions of conservation of the baryon number and the process going with overcoming of the minimum $Q$ value lead to the assumption about the production of $^6He$. For example, this process might just be $^6He + {}^4He + {}^1H + n + {}^2H$ ($Q \approx 49$ MeV). It can be presented as a charge exchange of one of the α-clusters to an unbound state $^4H$ and a subsequent capture of a neutron pair by another alpha cluster.

The results presented illustrate the research potential of relativistic nuclear dissociation for the investigation of various systems of lightest nuclei. These possibilities can be realized in nuclear experiments of a new generation which are remarkable for their high complexity and variety of detectors. The nuclear emulsion method possibilities are found to be highly valuable for motivation and planning of such experiments.

In conclusion, we should note an applied significance of the accumulation of data on the production probabilities and the compound of the products of relativistic fragmentation of $^{14}N$ nuclei. Bombarding of the upper layers of the Earth's atmosphere by relativistic protons of galaxy origin during the time of its existence could result in the production of light rare-earth elements $Li$, $Be$ and $B$ with a subsequent accumulation of them in the Earth's surface. Investigations of the $^{14}N$ fragmentation structure in an inverse kinematics can lead one to some important conclusions about isotopes reserves in the earth's crust.

# REFERENCES

1. Websites "The BECQUEREL Project" http://becquerel.jinr.ru/ and http://becquerel.lhe.jinr.ru.

2. *P. A. Rukoyatkin et al.* Beams of relativistic nuclear fragments at the nuclotron accelerator facility // Czech. J. of Phys, Suppl. C. - 2006. Vol. 56. - P. 379 -384.

3. *H. H. Heckman et al.* Fragmentation of $^4$He, $^{12}$C, $^{14}$N, and $^{16}$O nuclei in nuclear emulsion at 2.1 Gev / nucleon // Phys. Rev C – 1978. - Vol. 17. – P. 1735 - 1747.

4. *V. V. Belaga et al.,* Coherent dissociation $^{12}$C$\to$3$\alpha$ in lead-enreached emulsion at 4.5 GeV/c per nucleon // Phys. At.Nucl. - 1995. - Vol. 58. - P. 1905 - 1910.

5. *N. P. Andreeva et al.* Coherent Dissociation $^{16}$O $\to$ 4$\alpha$ in Photoemulsion at an Incident Momentum of 4.5 GeV/*c* per Nucleon // Phys. At. Nucl. - 1996. –Vol. 59. - P. 102 - 109.

6. *M. I. Adamovich et al.,* Investigation of clustering in light nuclei by means of relativistic multifragmentation processes // Phys. At. Nucl. - 2004. - Vol. 67. - P. 514 - 518.

7. *D. A. Artemenkov et al.*// arXiv: nucl-ex/0610023.

8. *D. A. Artemenkov, T. V. Shchedrina, R. Stanoeva, and P. I. Zarubin* Clustering features of *$^9$Be*, *$^{14}$N*, *$^7$Be*, and *$^8$B* nuclei in relativistic fragmentation // Conf. Proc. of International Symposium on Exotic Nuclei / Khanty-Mansiysk (Russia), 2007, Vol. 912, - P. 78 - 87; arXiv: 0704.0384.

9. *D. A. Artemenkov et al.* Features of the $^9$Be$\to$2He in an emulsion for an energy of 1.2 GeV per nucleon // Phys. At. Nucl. - 2007. – Vol. 70. -P. 1222 - 1226; arXiv:nucl-ex/0605018.

10. *N. G. Peresadko et al.* Fragmentation channels of relativistic $^7$Be nuclei in peripheral interactions // Phys. At. Nucl. - 2007. Vol. 70. - P. 1226 - 1230; arXiv: nucl-ex/0605014.

11. *R. Stanoeva et al.* Peripheral fragmentation of $^8$B nuclei in nuclear emulsion at an energy of 1.2 GeV per nucleon // Phys. At. Nucl. - 2007. Vol. 70. – P. 1216 - 1222; arXiv: nucl-ex/0605013.

12. *T. V. Shchedrina et al.* Peripheral interactions of relativistic $^{14}$N nuclei with emulsion nuclei // Phys. At. Nucl. - 2007. Vol. 70. -P. 1230 -1235; arXiv:nucl-ex/0605022.

13. *N. P. Andreeva et al.* Topology of "white stars" in the relativistic fragmentation of light nuclei // Phys. At. Nucl. - 2005. Vol. 68. – P. 455 – 466; arXiv:nucl-ex/0605015.

14. *M. I. Adamovich et al* Interactions of relativistic $^6$Li nuclei with photoemulsion nuclei // Phys. At. Nucl. - 1999. Vol. 62. – P. 1378 - 1387.